\documentclass[aps,prb,reprint,amsmath,amssymb,superscriptaddress,showpacs,floatfix,twocolumn,longbibliography]{revtex4-2}
\usepackage{graphicx}
\usepackage{dcolumn}
\usepackage{bm}
\usepackage[colorlinks, allcolors=blue]{hyperref}
\usepackage[usenames, dvipsnames]{color}
\usepackage[normalem]{ulem}
\usepackage{soul}
\usepackage[dvipsnames]{xcolor}

\begin{document}

\title{Superconductivity from interband coupling to ferroelectric quantum critical fluctuations in two dimensions}

\author{Sudip Kumar Saha}
\affiliation{Department of Physics, Bar Ilan University, Ramat Gan 5290002, Israel}

\author{Jonathan Ruhman}%
\affiliation{Department of Physics, Bar Ilan University, Ramat Gan 5290002, Israel}

\author{Avraham Klein}
\email{avraham.klein@biu.ac.il}
\affiliation{Department of Physics, Bar Ilan University, Ramat Gan 5290002, Israel}

\date{\today}

\begin{abstract}
Soft critical fluctuations associated with ferroelectric quantum phase transitions are 
typically transverse owing to their polar nature. This implies that the 
conventional density--density electron--phonon coupling to these modes is strongly suppressed, which is puzzling as a variety of materials exhibit enhanced superconductivity in the vicinity of ferroelectricity. An alternative coupling mechanism is an interband ``Stark''-like coupling that connects bands of opposite parity. 
In the limit where one of the bands is far in energy, these processes generate an effective quadratic (two-phonon) coupling. In contrast, when both bands lie close to the Fermi energy, the resulting interaction develops singular behavior due to the additional gapless electronic states, motivating a detailed study into the dynamics of this effective two-phonon coupling. To this end, we construct the quantum critical Eliashberg theory for a two-dimensional system across a wide 
range of interband gap magnitudes, near the quantum critical point.
We find that the critical temperature $T_c$ is strongly enhanced relative to conventional BCS expectations. This enhancement becomes more pronounced as the interband gap decreases.
In the large-gap limit, the pairing kernel acquires higher-order logarithmic contributions, leading to a parametrically enhanced $T_c$ governed by cubic and quadratic logarithmic terms. 
In the small-gap regime, the pairing scale exhibits a modified BCS-like form with an enhanced dependence on the inverse square root of the dimensionless coupling constant. 
The enhancement is due to the dynamics of the two-phonon pairing whose infrared cutoff is set by $T_c$, resulting in a significant enhancement of superconductivity compared to three-dimensional systems, where it is set by the Fermi energy.
Our results elucidate the unique dynamical properties of effective two-phonon interactions, and may be relevant to layered compounds like Td-MoTe$_2$ and doped SrTiO$_3$ membranes.
\end{abstract}
\maketitle
\section{\label{sec:intro}Introduction}

The conventional ferroelectric (FE) transition in an insulator typically arises from polar distortions that break inversion symmetry, resulting in a reversible bulk polarization. In metals, however, it is suppressed due to screening by itinerant electrons, which seems to rule out the existence of an FE order in the presence of gapless electronic excitations. 
Nevertheless, certain materials defy this notion by retaining broken inversion symmetry in the presence of dilute carrier concentration. For instance, doped SrTiO$_3$~\cite{Rischau2017SrCaTiO3} and KTaO$_3$~\cite{Zhang2023KTaO3} exhibit ferroelectricity under chemical substitution or strain, although ferroelectricity is absent in pristine samples due to strong quantum fluctuations. Other examples of polar metals include IV-VI semiconductors (PbTe, SnTe, GeSe)~\cite{Kobayashi1976SnTe, Guan2018GeSe} and bilayer transition metal dichalcogenides (TMDs)~\cite{Jindal2023}. They often exhibit superconductivity (SC) in the vicinity of a quantum critical point (QCP) separating the ordered and disordered states, along with anomalous transport properties. In many of these materials, the electronic band structure is significantly influenced by the combined effect of inversion symmetry and strong spin-orbit coupling (SOC) ~\cite{mattheiss1972, vandermarel2011,karsten2013,jonathan2020,maria1, maria2}. This indicates that in these so-called Quantum Ferroelectric Metals (QFEMs), there is a significant interaction between the correlated electrons and polar phonons.

The origin of SC in QFEMs is a subject of intense discussion across both theoretical and experimental platforms. The pairing mechanism has been attributed to the critical fluctuations of the polar phonon mode~\cite{edge2015quantum, wolfle2018superconductivity,maria2}, which is supported by experimental observations of the enhancement of $T_c$ when these systems are tuned towards QCP by doping or strain. However, theoretical attempts to justify this pairing mechanism must account for an important issue. 
In the ultra-dilute limit, when the electronic plasma frequency is smaller than the polar longitudinal optical (LO) modes, the soft polar fluctuations near the QCP become purely transverse optical (TO)~\cite{kadlec2009, roussev2003theory, yamanaka2000, jonathan2020, avi_qfem}. Namely, the long-range Coulomb interaction stiffens the LO phonons, leading to the LO-TO  splitting. Typically, the dominant mechanism for coupling phonons to electrons in metals arises from a density-density interaction, where fluctuations of the electronic charge density couple to the lattice deformation potential, proportional to the divergence of the ionic displacement.
Such a coupling to the transverse branch vanishes near the zone-center and is therefore irrelevant to the low-density limit of QFEMs~\cite{ruhman2019comment}. Hence, the nature of the underlying electron--phonon (el--ph) coupling that leads to pairing in QFEMs remains highly debated.

Alternative coupling mechanisms have been proposed. The first is the dynamical Rashba coupling, which is linear in the phonon displacement~\cite{kozii2015odd, ruhman2016superconductivity, kozii2019superconductivity, avi_qfem,maria1,maria2}.  
Another possible mechanism is the quadratic coupling to the soft mode, which was originally introduced by Ngai~\cite{ngai1974} and later reemerged as a potential source of pairing in QFEMs~\cite{vandermarel2019,feigelman_STO_1,volkov2022}. Recent work has further emphasized that two-boson couplings can qualitatively modify the low-energy critical theory itself, potentially driving strong-coupling behavior and novel critical phenomena near quantum critical points~\cite{NguyenKozii2025}.
Both linear and nonlinear mechanisms can lead to the emergence of SC near the QCP ~\cite{kozii2015odd, wang2016topological, maria1, maria2, vandermarel2019, kozii2019superconductivity, volkov2020multiband, volkov2022, avi_qfem, feigelman_STO_2, feigelman_STO_1, chaudhary2023superconductivity}. Doped SrTiO$_3$, the most celebrated QFEM, provides an ideal platform to test them. A recent investigation of ours ~\cite{smja2025} showed that the linear Rashba coupling plays a dominant role at optimal doping and in the overdoped regime. 
However, because the Rashba coupling originates from bond distortions rather than charge density fluctuations, it becomes weak in the low-density limit. Thus, the nonlinear coupling dominates in this limit~\cite{vandermarel2019, feigelman_STO_1, volkov2022}. Moreover, the nonlinear correction is necessary near the top of the superconducting dome to account for the shift of the maximum $T_c$ from the QCP into the ordered state ~\cite{smja2025}.

Most studies focus on the single-band model, where the nonlinear coupling is introduced on a phenomenological basis (see Ref.~\cite{vandermarel2019} for an exception).  The single-band approximation is justified when the density of states (DOS) of the lowest band greatly exceeds that of higher bands, as in dilute SrTiO$_3$, where interband interactions provide only a small correction to the dominant intraband pairing~\cite{wolfle_balatsky2013}.

A detailed analysis of the multiband origin of the nonlinear coupling to the FE QCP in QFEMs is, however, essential. When several bands lie close in energy with comparable DOS, interband processes can strongly modify SC, leading to deviations between theory and experiment. Microscopically, the leading mechanism for a quadratic el--ph coupling arises from virtual interband transitions: an electron temporarily hops to an orbital of opposite parity by emitting or absorbing a TO phonon and subsequently returns via the reverse process, generating an effective two-phonon interaction~\cite{vandermarel2019}.  
Optical spectroscopy in SrTiO$_3$~\cite{vandermarel2019} has identified such transitions between O $2p$ and Ti $3d$ orbitals as a relevant nonlinear pairing channel at ultralow densities. Unlike phenomenological approaches, a microscopic treatment also enables the inclusion of retardation effects that may become significant near the QCP.

\begin{figure}[]
\includegraphics[width=\hsize]{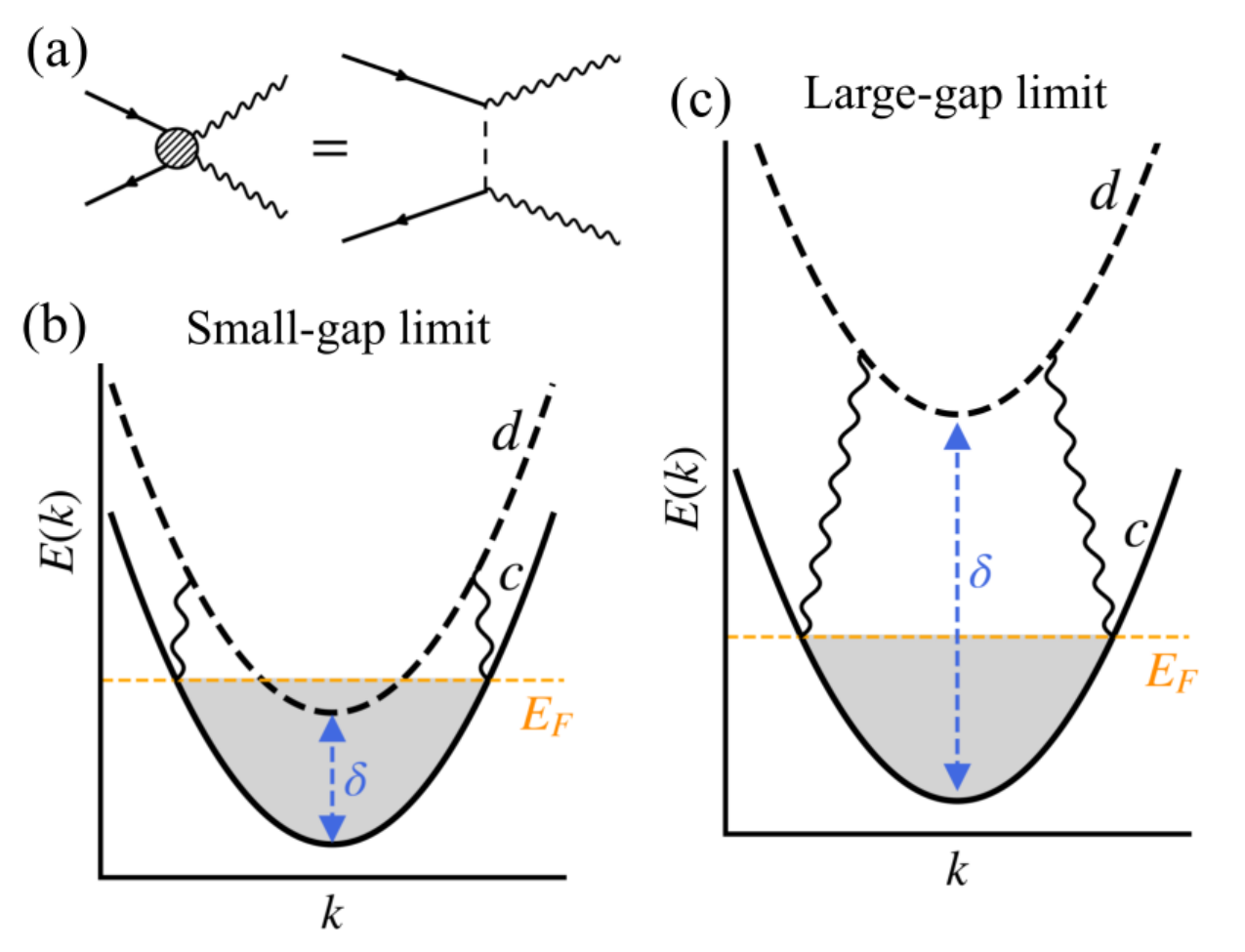}
\caption{\label{model_diag} The schematic diagram of the two-band model and the interaction via the two-phonon mechanism. 
(a) The diagrammatic representation of the two-phonon process. The solid and dashed straight lines represent fermions of lower and upper bands, respectively. The wavy lines represent TO phonons.
(b,c) The schematic illustration of the interband interactions mediated by TO phonons in the (b) small-gap limit and (c) large-gap limit. The lower- and higher-energy parabolic bands are shown by solid and dashed lines, respectively, separated by a band gap $\delta$. The dashed horizontal line marks the Fermi energy, $E_F$, and the shaded region denotes the occupied electronic states below $E_F$. 
}
\end{figure}

In order to address this issue, in this paper we study a two-dimensional (2D) model of QFEMs where the coupling to the critical FE  fluctuations emerges from the exchange of TO phonons between two electronic bands, as illustrated schematically in Fig.~\ref{model_diag}. This process gives rise to an effective quadratic coupling (Fig.~\ref{model_diag}(a)), even though the underlying el--ph interaction is assumed to be linear. We employ a strong-coupling (Eliashberg) pairing analysis to investigate pairing in this system at the FE QCP.
Our model allows us to vary the bandgap $\delta$ while remaining tuned to the QCP, and see how the pairing evolves as one goes from large to small gaps (see Fig.~\ref{model_diag}(b) and \ref{model_diag}(c)). Because we are interested in the theoretical insights derived from the microscopic mechanism, we choose to study a 2D system, where quantum fluctuations are strong, rather than the three-dimensional (3D) mechanism studied previously in the context of e.g. SrTiO$_3$. And, indeed, we find that the behavior in 2D system is significantly more interesting than the 3D case. 
In the latter case, at ultralow densities, the Landau damping is suppressed. Consequently, the pairing attraction exhibits a logarithmic divergence, which is cut off by the small Fermi surface  ~\cite{feigelman_STO_1, volkov2022, smja2025}. At higher densities, the Fermi surface grows and the nonlinear contribution becomes negligible. 
However, in a 2D system, the nonlinear mechanism originating from the interband interaction, yields a stronger attraction whose IR cutoff is $T_c$, despite the bosons being underdamped and the Fermi surface being large. Hence, $T_c$ is parametrically enhanced compared to BCS theory, and it is further enhanced when one goes to small gap regime.

{In this work, we specifically focused on the ferroelectric case because (a) as we discussed, the linear coupling is expected to vanish at low density, and (b) SrTiO$_3$ is a concrete model system that has been shown to host a robust two-phonon coupling. Indeed, our results are directly relevant to doped SrTiO$_3$ membranes ~\cite{Eom_membrane_2015, Hwang_membrane_2020, Hwang_membrane_2021, Jespersen_membrane_2022, Lee_membrane_2025}. However, we emphasize that our model and results should be relevant to any type of quantum-critical point in which the linear coupling to electrons is weak at low densities due to symmetry considerations, since the quadratic coupling is a scalar.}


\section{\label{sec2}Model }

To capture the essential physics of QFEMs, we assume a minimal model that includes itinerant fermions, TO phonons, and their coupling. 
The Hamiltonian describing the low-energy electrons of a two-band system is
\begin{equation}
 \begin{aligned}
H_{FL}= \sum_{\bm{k};\sigma} \left\{  c_{\bm{k},\sigma}^\dagger   \left( \epsilon_k  -\mu \right)  c_{\bm{k},\sigma} +    d_{\bm{k},\sigma}^\dagger   \left[ \epsilon_k  -\left(\mu-\delta\right) \right]  d_{\bm{k},\sigma}\right\} \, .
\label{eq:FL_ham}
 \end{aligned}
\end{equation}
Here, we consider a 2D electronic system with momentum $\bm{k} = (k_x,k_y)$. The fermionic flavors of the lower and upper bands are represented by the annihilation (creation) operators $c$ ($c^\dagger$) and $d$ ($d^\dagger$), respectively, and $\sigma$ labels their spin degree of freedom. We assume that both bands are parabolic and have the same mass $m$, with dispersions $\epsilon_{\bm{k}}^c =\epsilon_{\bm{k}}=\bm{k}^2/2m$ and $\epsilon_{\bm{k}}^d=\epsilon_{\bm{k}}+\delta$, where $\bm{k}^2 = k_x^2+k_y^2$. The lower band is assumed to lie near the Fermi level $E_F$ ($\epsilon_{\bm{k}} \approx E_F$). The chemical potential and energy difference between the bands are denoted by $\mu$ and  $\delta$, respectively.

Another essential building block of our minimal model is the 2D TO phonon, described by the bosonic propagator, 
 \begin{equation}
D(q)=D_0 \left(r+\vert \bm{q} \vert^2 a^2+\frac{q_0^2a^2}{c^2}\right)^{-1} \,.
\label{eq:boson_prop_ham}
 \end{equation}
Here, $q=(q_0,\bm{q})$ represents the bosonic three-vector, where $\bm{q} = (q_x, q_y)$ is the two-dimensional (2D) bosonic momentum and $q_0$ denotes the bosonic Matsubara frequency. The lattice constant and the TO phonon velocity are denoted by $a$ and $c$, respectively.  The constant $D_0$ has the unit of inverse energy.  The parameter $r=a^2/\xi^2$ measures the distance from the QCP, with $\xi$ being the correlation length that diverges at the QCP.

Finally, we assume an interband interaction mediated by the TO phonon mode, leading to a two-phonon process where the two contributions are Hermitian conjugates of each other. The corresponding Hamiltonian is given by 
\begin{equation}
H_{int}=g \sum_{\bm{k},\bm{q};\sigma} \eta(\bm{q}) \left( c_{\bm{k}+\frac{\bm{q}}{2} ,\sigma}^\dagger  d_{\bm{k}-\frac{\bm{q}}{2},\sigma}  +h.c.  \right) \, ,
\label{eq:int_ham}
 \end{equation}
 where $\bm{\eta}$ is the transverse component of the dimensionless phonon displacement $\bm{u}$ such that
\begin{equation}
 \begin{aligned}
	\bm{\eta}(\bm{q})=
 \hat{P}(\hat{q})\cdot\bm{u}(\bm{q})\, , \qquad\quad P_{ij}(\hat{q})=\delta_{ij}-\hat{q}_i \hat{q}_j \, .
	 \label{eq:projection_define}
 \end{aligned}
\end{equation}
Here, $P_{ij}$ and $\hat{P}$ are the components and the matrix form of the projection operator on the transverse sector, respectively.


\section{\label{sec3}Self Energy}
In this section, we investigate the dynamics of the two-band QFEM model at the QCP, captured by the self-energies of bosons and fermions. The main results for the self-energies at one-loop order are presented in this section, while the full derivation is provided in the Appendix.

\begin{figure*}[]
\includegraphics[width=\hsize]{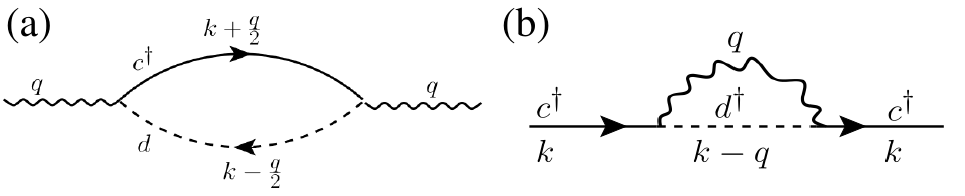}
\caption{\label{self_energy_diag}
The diagram of the self-energy. Panel (a) represents the one-loop polarization bubble and panel (b) represents the fermionic self-energy. The wavy lines represent the boson propagator. The solid and dashed straight lines represent Green’s functions corresponding to fermionic flavors of lower and upper bands, respectively. }
\end{figure*}

The bosonic self-energy is derived from the polarization bubble, shown diagrammatically in Fig.~\ref{self_energy_diag}(a), and is given by the following expression
\begin{equation}
\Pi_{d^\dagger c/c^\dagger d}(q) =\bar{g} a^2 T \sum_k G^{d/c}(k+q/2) G^{c/d}(k-q/2) \, .
\label{eq:bubble}
 \end{equation}
Here, the propagators $G^c(k)$ and $G^d(k)$ correspond to the fermionic flavors of low-energy and high-energy bands, respectively. The fermionic three-vector is defined as $k\equiv(k_0,\bm{k})$, where $ \bm{k}$ is the two-dimensional (2D) momentum and $k_0$ is the Matsubara frequency. The effective fermion-boson interaction is given by $\bar{g}= g^2 D_0$. 
At zero-temperature limit of an infinite system, the sum over Matsubara frequencies and momenta is replaced by integrations as $T\sum_k \rightarrow \int  dk_0/2\pi \int d^2k/(2\pi)^2 $. The integration over $2D$ momenum is then decomposed into energy and angular components (see Appendix).  After performing the integration over frequency and energy, we obtain
\begin{equation}
 \begin{aligned}
\Pi_{d^\dagger c}(q) =\bar{g} \nu_F \int_0^{2\pi}\left[1+\frac{  iq_0/v_F \vert \bm{q} \vert}{ \cos\theta_k - i(q_0+i\delta)/v_F \vert \bm{q} \vert }  \right]  \, \frac{d\theta_k}{2\pi} \, .
 \label{eq:bubble_int1}
 \end{aligned}
 \end{equation}
Here, $\nu_F=k_F a^2/2\pi v_F=m^* a^2/2\pi$ is the 2D DOS per spin at the Fermi level, where $v_F$ denotes the Fermi velocity. It is now straightforward to examine two limiting cases.
In the large band gap limit ($\delta \gg v_F \vert \bm{q} \vert \gg q_0 $), the angular integration in Eq.~\eqref{eq:bubble_int1} yields 
\begin{equation}
 \begin{aligned}
\Pi_{d^\dagger c}(q) =  \bar{g}\nu_F \left[ 1+i \frac{q_0}{\delta} -\frac{q_0^2}{\delta^2} +\mathcal{O}\left(\frac{q_0^3}{\delta^3} \right) \right]\,.
 \label{eq:bubble_int2}
 \end{aligned}
 \end{equation}
To compute the bosonic self-energy in the two-band model, we must consider Eq.~\eqref{eq:bubble_int2} together with its complex conjugate, 
\begin{equation}
\Pi(q)=\Pi_{d^\dagger c}(q)+
\Pi_{c^\dagger d}(q)=\bar{g}\nu_F \left[ 1-\frac{q_0^2}{\delta^2}  \right]\, ,
 \label{eq:boson_self_lm1}
 \end{equation}
where we have neglected the higher-order terms. 
The static contribution in this expression renormalizes the phonon mode energy as $r \rightarrow r-\bar{g}\nu_F$. 
At the QCP ($r = 0$), the bosonic propagator $D(q)$ takes the form
  \begin{equation}
  D(q)=D_0 \left(\vert \bm{q} \vert^2 a^2+\frac{ q_0^2 a^2} {\tilde{c}^2}  \right)^{-1}  \, ,
\label{eq:boson_propagator_lm1}
 \end{equation}
resulting in an underdamped boson. Here, we have introduced a new parameter, the renormalized phonon velocity $\tilde{c}$, which is related to $c$ as $1/\tilde{c}^2=1/c^2+\bar{g} \nu_F/\delta^2 a^2$.
 In contrast, in the small band gap limit ($q_0 \ll \delta \ll  v_F \vert \bm{q} \vert$), the bosons become overdamped because the self-energy acquires a Landau damping term. At the QCP, $D(q)$ has the form
  \begin{equation}
  D(q)=D_0 \left( \vert \bm{q} \vert^2 a^2+   \frac{\bar{g} \nu_F \vert q_0 \vert} {v_F \vert \bm{q} \vert} \right)^{-1}  \, .
\label{eq:boson_propagator_lm2}
 \end{equation}

Next, we compute the fermionic self-energy $\Sigma(k)$, shown diagrammatically in Fig.~\ref{self_energy_diag}(b), which assumes the form
\begin{equation}
\Sigma(k_0)=\frac{\bar{g}a^2}{(2\pi)^3}\iint G^d(k-q)D(q)  \,d^2q  \, dq_0\, .
\label{eq:fermion_self}
 \end{equation}
 After factorizing the integration over the 2D momentum $d^2q$ and carrying out the angular integration, Eq.~\eqref{eq:fermion_self} reduces to
\begin{equation}
 \begin{aligned}
\Sigma(k_0)&= \frac{-i\bar{g}a^2}{(2\pi)^2} \int dq_0 \int dq \, q  D(q)   \\
& \times\left(\frac{\text{sgn}(k_0-q_0)}{\sqrt{\left(v_F \vert \bm{q} \vert \right)^2+ \left(k_0-q_0+i\delta \right)^2  }}\right) \,.
\label{eq:fermion_self1}
 \end{aligned}
 \end{equation}
 In the large gap limit ($\delta \gg v_F \vert \bm{q} \vert \gg \vert k_0-q_0 \vert$), this expression yields a static contribution $ \Sigma \approx -\bar{g}a^2\Lambda \tilde{c}/4\pi \delta $, which shifts the chemical potential. Here, $\Lambda$ denotes the UV momentum cutoff. On the other hand, in the small gap ($v_F \vert \bm{q} \vert \gg \delta \gg \vert k_0-q_0 \vert$), small frequency limit, it simplifies to a frequency-dependent form $\Sigma(k_0)\approx  - (i\bar{g}a^2/2\pi^2) (k_0/\delta)$. This linear dependence of self-energy on frequency is characteristic of Fermi liquid (FL) behavior.


\section{\label{sec4} Pairing}

\begin{figure}[t]
\includegraphics[width=\hsize]{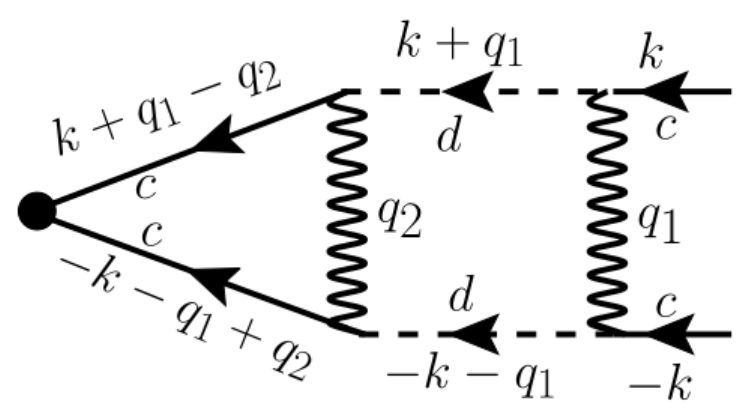}
\caption{\label{vertex_diag}
The diagrammatic representation of the pairing vertex. The wavy lines represent the boson propagator. The solid and dashed straight lines represent Green’s functions corresponding to fermionic flavors of lower and upper bands, respectively. }
\end{figure}

We now turn our attention to the superconductivity at the QCP, where FE fluctuations are the strongest. The linearized equation for the pairing vertex, shown diagrammatically in Fig.~\ref{vertex_diag}, is given by 
\begin{equation}
 \begin{aligned}
& \qquad \phi(k_0) =\bar{g}^2 a^4 T^2 \sum_{q_{1,0}}\sum_{q_{2,0}}\int\frac{d^2q_1}{(2\pi)^2} \int\frac{d^2q_2}{(2\pi)^2} \\
& G^c(-k-q_1+q_2) \phi(k_0+q_{1,0}-q_{2,0})  G^c(k+q_1-q_2)D(q_2) \\ 
& \qquad \times G^d(k+q_1)G^d(-k-q_1)D(q_1) \,.
 \label{eq:vertex1}
 \end{aligned}
\end{equation}
where $q_{1,0}$ and $q_{2,0}$ denote the temporal (Matsubara frequency) part of the vectors $q_1$ and $q_2$, respectively. 
Here, the leading pairing instability is assumed to come from singlet pairing. This expression involves summations over bosonic Matsubara frequencies $q_{1,0}$ and $q_{2,0}$, and integrations over bosonic momenta $q_1$ and $q_2$. In the following subsections, we analyze this expression in both large-gap and small-gap limits and determine the dependence of $T_c$ on the band gap. 

\begin{figure*}[t]
\includegraphics[width=0.8\hsize]{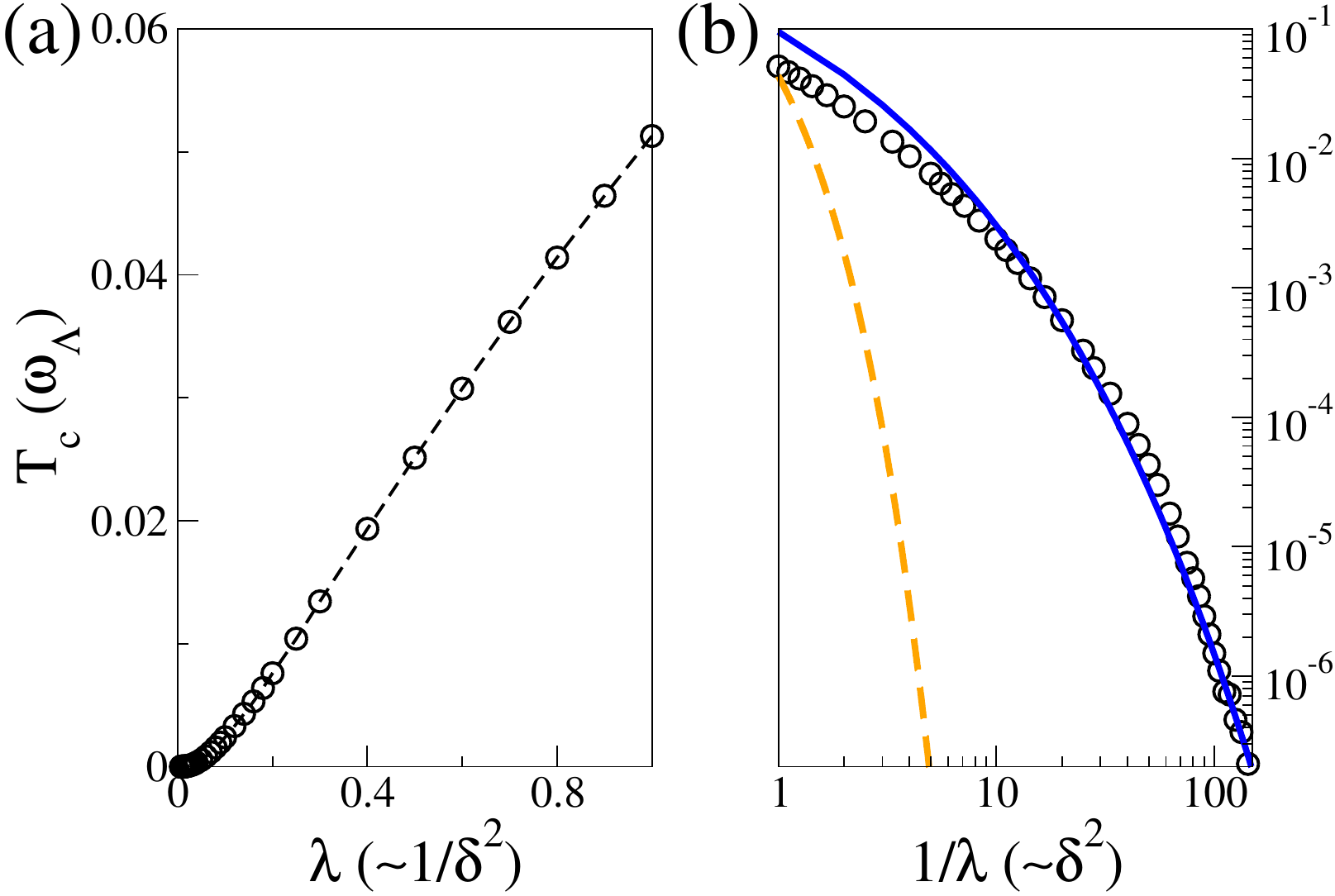}
\caption{\label{tc_num_soln} $T_c$ (in unit of $\omega_\Lambda$), derived from Eq.~\eqref{eq:vertex1_eigen_lm1}, as a function of coupling constant, $\lambda$ in panel (a) and $1/\lambda$ in panel (b). 
The numerical solutions are shown as circles, with a black dashed line added as a guide to the eye. 
The solid blue line presents the solution of Eq.~\eqref{eq:vertex1_freqsum1_lm1}. The solution of BCS expression, $(\lambda/\pi)\log{(\omega_\Lambda/T_c^{BCS})}=1$, is depicted by the yellow dashed line.
}
\end{figure*}

\subsection{\label{sec4_1}  Large-gap limit  }
We first analyze Eq.~\eqref{eq:vertex1} in the large-gap limit ($\delta \gg v_F \vert \bm{q}_1 \vert \gg q_{1,0} $ and $\delta \gg v_F \vert \bm{q}_2 \vert \gg q_{2,0} $), where phonons are underdamped and the fermionic self-energy merely renormalizes the chemical potential through a static contribution, as discussed in Section~\ref{sec3}.
Upon integrating over momenta, Eq.~\eqref{eq:vertex1} reduces to
\begin{equation}
 \begin{aligned}
\phi(\tilde\omega_m)= \lambda T^2 &\sum_{m^\prime \ne m^{\prime \prime} }\sum_{  m^{\prime \prime} \ne m  }  \frac{\phi (\tilde\omega_{m^\prime})}{\vert \tilde\omega_{m^\prime} \vert}  \frac{1 }{\vert \tilde\omega_{m^{\prime \prime}} -\tilde\omega_m{^\prime} \vert} \\
&\times  \log{\left(1+\frac{\omega_\Lambda^2}{   \left( \tilde\omega_{m^{\prime \prime}} -\tilde\omega_m  \right)^2 }\right)}\, ,
 \label{eqmn:vertex1_freq1_lm1}
 \end{aligned}
\end{equation}
where we denote $\tilde \omega_m = k_0$, $\tilde \omega_{m^{\prime}}=k_0+q_{1,0}-q_{2,0}$ and $\omega_{m^{\prime \prime}}=k_0 + q_{1,0}$ (See the Appendix \ref{app2} for a detailed calculation). Therefore, $\tilde\omega_{m^\prime}=(2m^\prime+1)\pi T$ denotes fermionic Matsubara frequencies, while $ \tilde\omega_{m^{\prime \prime}} -\tilde\omega_m{^\prime}=2\pi T (m^{\prime \prime}-m^{\prime }) $ and $\tilde\omega_{m^{\prime \prime}} -\tilde\omega_m=2\pi T (m^{\prime \prime}-m ) $ correspond to bosonic Matsubara frequencies at temperature $T$. The indices $m$, $m{^\prime}$ and $m^{\prime \prime}$ represent integer numbers.  
The dimensionless parameter $\lambda$ represents the effective coupling constant and is given by 
\begin{equation}
    \lambda =\frac{\tilde{c}  }{16\pi v_F }\left(\frac{\bar{g}}{\delta}\right)^2 \,.
\end{equation}
We have also defined $\omega_\Lambda=\Lambda \tilde{c}$ with $\Lambda$ being the UV momentum cutoff. 
Notably, the diverging terms at zero frequency transfer, which account for the thermal fluctuations, have been excluded. Typically, these are regularized by the boson thermal mass. Here, for computational convenience, we just remove them, as we are more interested in the qualitative trends of $T_c$ than in a quantitatively exact computation.

As usual, the linearized gap equation can be recast as a zero-eigenvalue problem,
\begin{equation}
\sum_{m^\prime=0}^{M}
K_{m,m^\prime}(\tilde T)\,
\phi(\tilde\omega_{m^\prime})
=0 \, ,
\label{eq:vertex1_eigen_lm1}
\end{equation}
with
\[
K_{m,m^\prime}
=
\sum_{\substack{m^{\prime\prime}=0\\
m^{\prime\prime}\neq m,\;m^\prime}}^{M}
R_{m,m^\prime,m^{\prime\prime}}(\tilde T)
-\delta_{m,m^\prime} \,,
\]
where
\begin{widetext}
\begin{equation}
\begin{aligned}
R_{m,m^\prime,m^{\prime\prime}}
=
\frac{\lambda}{2\pi^2(2m^\prime+1)}
\left(
\frac{1}{|m^{\prime\prime}-m^\prime|}
+
\frac{1}{|m^{\prime\prime}+m^\prime|}
\right)
\log\!\left[
\left(
1+\frac{1}{4\pi^2\tilde T^2(m^{\prime\prime}-m)^2}
\right)
\left(
1+\frac{1}{4\pi^2\tilde T^2(m^{\prime\prime}+m)^2}
\right)
\right] \, .
\label{eq:vertex1_eigen1_lm1}
\end{aligned}
\end{equation}
\end{widetext}
Here, $\tilde T=T/\omega_\Lambda$. We have also performed the variable shifts
$-\tilde\omega_{m^\prime}\to\tilde\omega_{m^\prime}$ and
$-\tilde\omega_{m^{\prime\prime}}\to\tilde\omega_{m^{\prime\prime}}$,
and truncated all frequency sums at $m=M$.
The kernel $K_{m,m^\prime}(\tilde T)$ is therefore an
$(M+1)\times(M+1)$ matrix, while $\phi$ is an $(M+1)$-component column vector.
The transition temperature is determined by the condition that the kernel develops a zero eigenvalue, equivalently
\[
\det K(\tilde T)=0 \,.
\]

In what follows, we set the frequency cutoff to $M=2000$ and solve
Eq.~\eqref{eq:vertex1_eigen_lm1} numerically to determine $T_c$.
This cutoff is sufficient for $T_c$ to be only weakly dependent on $M$ (see Appendix~\ref{app2} for convergence tests).
The resulting $T_c$ values in the large-gap limit are shown in
Fig.~\ref{tc_num_soln}(a) as black circles. As expected, $T_c$ increases with the
coupling strength $\lambda$. In the weak-coupling limit, however, we find
$T_c\sim \exp(-1/\sqrt{\lambda})$, which is parametrically larger than the
standard BCS result. This enhancement is illustrated in Fig.~\ref{tc_num_soln}(b),
where we plot $T_c$ as a function of $1/\lambda$ on a log-log scale and compare
it with the BCS scaling $T_c^{\rm BCS}\propto \exp(-1/\lambda)$, shown as a
yellow dashed line.

To identify the origin of this enhancement, we analyze
Eq.~\eqref{eqmn:vertex1_freq1_lm1} analytically. Using $2\pi T_c$ as the lower
frequency cutoff and integrating over frequencies, we obtain
\begin{equation}
\frac{\lambda}{2\pi^2}
\left[
\frac{2}{3}
\left(
\log\frac{\omega_\Lambda}{T_c}
\right)^3
+
2
\left(
\log\frac{\omega_\Lambda}{T_c}
\right)^2
\right]
=1 \,.
\label{eq:vertex1_freqsum1_lm1}
\end{equation}
The derivation is given in Appendix~\ref{app2}. In contrast to the BCS case,
where the instability is controlled by a term linear in $\log T_c$, here the
quadratic and cubic logarithmic terms in
Eq.~\eqref{eq:vertex1_freqsum1_lm1} produce the parametrically enhanced scaling
of $T_c$. The solution of Eq.~\eqref{eq:vertex1_freqsum1_lm1}, shown as a blue
solid line in Fig.~\ref{tc_num_soln}, agrees well with the full numerical
solution.
 

\subsection{\label{sec4_2}  Small-gap limit  }
Next, we turn our attention to the small-gap limit ($v_F \vert \bm{q}_1 \vert \gg \delta \gg  q_{1,0} $ and $v_F \vert \bm{q}_2 \vert \gg \delta \gg  q_{2,0} $).  
As established in Section~\ref{sec3}, our two-band model in this limit features overdamped bosons and FL behavior. However, the Landau damping term is suppressed here because  $\delta$ acts as an IR cutoff of momentum in the bosonic propagators $D(q_1)$ and $D(q_2)$ in Eq.~\eqref{eq:vertex1}. This enables a straightforward analytical derivation of $T_c$, obviating the need for a numerical approach. The main results are presented below, with the detailed derivation provided in the Appendix \ref{app4}.

After integrating out the momenta in Eq.~\eqref{eq:vertex1}, we obtain
\begin{equation}
 \begin{aligned}
 \phi(k_0)  =\lambda_s T^2 & \sum_{\substack{q_{1,0}; \\ \vert q_{1,0} \vert \ne \vert k_0 \vert   }}\sum_{\substack{q_{2,0}; \\ \vert q_{2,0} \vert \ne \\ \vert k_0 +q_{1,0} \vert   }} \phi(k_0+q_{1,0}-q_{2,0}) \\
 & \times\frac{1}{\vert k_0+q_{1,0}-q_{2,0} \vert}  \frac{1}{\vert k_0+q_{1,0} \vert} \, ,
 \label{eq:vertex1_freq_lm2}
 \end{aligned}
\end{equation}
where the effective coupling constant is given by
\[ \lambda_s \approx \bar{g}^2 /4\pi \delta^2 \]
(exact expression can be found in Appendix \ref{app4}).
After converting the frequency summations to integrations and subsequently integrating out the frequencies, Eq.~\eqref{eq:vertex1_freq_lm2} reduces to  
\begin{equation}
 \begin{aligned}
\frac{\lambda_s}{\pi^2}   \left[ \log{\frac{\Lambda_\Omega}{T_c}} +\frac{1}{2} \left(  \log{\frac{\Lambda_\Omega}{T_c}} \right)^2   \right] = 1 \, ,
 \label{eq:vertex1_freqsum2_lm2}
 \end{aligned}
\end{equation}
where $\Lambda_\Omega$ denotes the upper frequency cutoff.
We now introduce a new variable $x(T_c)=\log{(\Lambda_\Omega/T_c)}$, which leads to a quadratic equation $x^2+2x-2\pi^2/\lambda_s=0$. 
Its solutions are $x(T_c)=m_\pm=-1\pm \sqrt{1+ 2\pi^2/\lambda_s }\approx -1\pm \sqrt{ 2\pi^2/\lambda_s }$. 
Since $T_c$ must be smaller than $\Lambda_\Omega$ (i.e., $T_c/\Lambda_\Omega<1$), only the solution $x=m_+$ yields a physically meaningful $T_c$:  $T_c=\Lambda_\Omega e^{-m_+}$. Thus, $T_c$ takes a BCS-like form, $T_c=e \Lambda_\Omega  \exp{(-\sqrt{2\pi^2/\lambda_s} )}$, but is parametrically larger than $T_c^{\text{BCS}}$. This enhancement is a consequence of the dependence of $T_c$ on $\sqrt{1/\lambda_s}$, which originates from the squared logarithmic term in Eq.~\eqref{eq:vertex1_freqsum2_lm2}.

 Taken together, the large- and small-gap limits demonstrate that interband coupling to critical ferroelectric fluctuations generates superconductivity that is consistently stronger than expected from conventional BCS theory. While the microscopic origin of the enhancement differs in the two regimes, both arise from the singular structure of the pairing kernel induced by multiband processes near the quantum critical point.
 
\section{Conclusions}

In this work, we investigated superconductivity mediated by critical
ferroelectric fluctuations in a two-band QFEM.
Starting from a microscopic model in which TO phonons
couple linearly to electrons through interband transitions, we analyzed
the resulting pairing instability within Eliashberg theory at the
FE QCP. This framework naturally generates
an effective two-phonon pairing interaction while retaining the full
multiband structure and retardation effects of the underlying el--ph coupling.

Our central result is that interband processes parametrically enhance
SC compared to conventional BCS expectations. In the
large-gap regime, where the upper band is only accessed virtually, the
system reduces to an effective two-phonon pairing problem. The pairing
kernel acquires singular logarithmic contributions that lead to cubic
and quadratic powers of $\log(\omega_\Lambda/T_c)$ in the gap equation.
As a consequence, the transition temperature exhibits substantial
enhancement relative to the standard BCS form
$\exp(-1/\lambda)$.

We further showed that the enhancement persists in the small-gap
regime, where low-energy states from both bands actively participate in
the pairing process. In this limit, the critical theory is characterized
by overdamped bosonic fluctuations and an FL fermionic
self-energy. Nevertheless, the interband gap provides an infrared
cutoff that suppresses the singular effects of Landau damping in the
pairing channel. The resulting gap equation again contains higher-order
logarithmic contributions, yielding a transition temperature that scales as $T_c \sim \exp(-1/\sqrt{\lambda_s})$, thereby retaining  a parametrically enhanced dependence on the coupling strength.

An important aspect of our analysis is the role of dimensionality.
Unlike previous studies of nonlinear coupling (Ngai) mechanisms in 3D systems ~\cite{vandermarel2019,feigelman_STO_1,volkov2022}, where the pairing attraction is often limited by the size of the Fermi surface, the 2D system considered here exhibits a much stronger instability. The infrared cutoff of the pairing kernel is set directly by the superconducting scale itself, allowing critical FE fluctuations to generate a robust attractive interaction over a wide range of parameters. 
More broadly, our results establish that nonlinear el--ph
coupling generated by interband transitions can provide an efficient
route toward SC in QFEMs. The mechanism is particularly effective near a FE QCP, where soft TO phonons produce singular
pairing interactions even when conventional density--density coupling is suppressed. These findings offer a microscopic perspective on the
interplay between multiband physics, FE criticality, and
SC, and may be relevant to materials such as doped
SrTiO$_3$~\cite{Rischau2017SrCaTiO3}, KTaO$_3$~\cite{Zhang2023KTaO3}, IV--VI semiconductors~\cite{Kobayashi1976SnTe, Guan2018GeSe}, and other polar metals in which multiple bands participate in the low-energy electronic structure.

Several extensions of the present work remain open. In particular, it
would be interesting to investigate the evolution away from criticality,
the competition and cooperation between interband two-phonon and
Rashba-type coupling mechanisms, and the role of realistic multiband
electronic structures. Beyond these questions, the present model offers
a promising platform for sign-problem-free quantum Monte Carlo
simulations, enabling controlled studies of critical fluctuations and
nonperturbative effects beyond Eliashberg theory. Such calculations
would provide an important benchmark for the analytical results
presented here. We hope that the framework developed in this work will
serve as a useful starting point for addressing these questions in
future studies.

\section{Acknowledgments}
We thank Rafael Fernandes and Yuxi Zhang for helpful discussions. J.R. acknowledges funding by the Simons Collaboration on “New Frontiers in Superconductivity” and ISF
under grant No. 915/24.

\newpage

\appendix
\onecolumngrid
\section{\label{app1} Self-energies}
In this Appendix, we discuss the details of the derivation of bosonic and fermionic self-energies within the Eliashberg approximation. We compute the bosonic self-energy from the one-loop fermionic bubble, which assumes the form 
 \begin{equation}
 \begin{aligned}
\Pi_{d^\dagger c}(q) &=\bar{g} a^2 T \, \sum_k G^d(k+q/2) G^c(k-q/2)  \\
&=\bar{g} a^2\int \left(\frac{1}{(k_0+q_0/2)+i\epsilon^d_{\bm{k}+\bm{q}/2}}\right) \left( \frac{1}{(k_0-q_0/2)+i\epsilon^c_{\bm{k}-\bm{q}/2}} \right)  \,  \frac{dk_0}{2\pi} \, \frac{d^2k}{(2\pi)^2}     \\
&=\bar{g} \nu_F \int \frac{ \left[\theta(-\epsilon^c_{\bm{k}})-\theta(-\epsilon^d_{\bm{k}+\bm{q}})\right]}{ v_F \vert \bm{q} \vert \cos\theta +  \delta -i q_0}    \, d\epsilon_{\bm{k}} \, \frac{d\theta_k}{2\pi}    \\
&=\bar{g} \nu_F \int  \left[1+\frac{  \frac{iq_0}{v_F \vert \bm{q} \vert}}{ \cos\theta - \frac{i(q_0+i\delta)}{v_F \vert \bm{q} \vert} }  \right]  \, \frac{d\theta_k}{2\pi} \\
&=\bar{g} \nu_F \left[1 - \frac{\vert q_0 \vert}{\sqrt{(v_F \vert \bm{q} \vert)^2+(q_0+i\delta)^2}}     \right] \\
&\approx\bar{g} \nu_F \left[1 - \frac{\vert q_0 \vert}{\sqrt{ -\delta^2+2 i q_0 \delta}}     \right]  \\
&= \bar{g}\nu_F \left[ 1+i \frac{q_0}{\delta} -\frac{q_0^2}{\delta^2}  +\mathcal{O}\left(\frac{q_0^3}{\delta^3} \right)   \right] \,.
 \label{eqapp:bubble}
 \end{aligned}
 \end{equation}
Thus we obtain Eq.~\eqref{eq:bubble_int1} and Eq.~\eqref{eq:bubble_int2} in the main text.
In the derivation of Eq. \eqref{eqapp:bubble}, we  replaced the sum over Matsubara frequencies and momenta by integration and decomposed the integration over 2D momenum into energy and angular integration,
\begin{equation}
 \begin{aligned}
\bar{g} a^2  \frac{d^2k}{\left(2\pi\right)^2}= \frac{\bar{g}}{2\pi}  \left(\frac{m^* a^2}{2\pi}\right)  \left(\frac{k\, dk}{m^*}\right) \, d\theta_k  =\frac{\bar{g}}{2\pi}  \nu_F d\epsilon_{\bm{k}} \, d\theta_k  \, ,
 \label{eq:pi_integration_decompose_2D}
  \end{aligned}
 \end{equation}
At the small $|\bm{q}|$ limit, the energies of the lower (upper) band is assumed to take the form $\epsilon^{c}_{\bm{k}+\bm{q}}\sim \epsilon_{\bm{k}}+v_F \vert \bm{q} \vert \cos{\theta_k}$  ($\epsilon^{d}_{\bm{k}+\bm{q}}\sim \epsilon_{\bm{k}}+\delta+v_F \vert \bm{q} \vert  \cos{\theta_k}$), where $\epsilon_{\bm{k}} \approx \epsilon_F \approx 0$. In the sixth step of Eq.~\eqref{eqapp:bubble}, we have used the large band gap approximation ($\delta \gg v_F \vert \bm{q} \vert \gg q_0 $). Similarly, we obtain the complex conjugate of the polarization bubble 
\begin{equation}
 \begin{aligned}
\Pi_{c^\dagger d}(q) =  \bar{g}\nu_F \left[ 1-i \frac{q_0}{\delta} -\frac{q_0^2}{\delta^2} -\mathcal{O}\left(\frac{q_0^3}{\delta^3} \right)  \right] \, .
 \label{eqapp:bubble_cc}
 \end{aligned}
 \end{equation}
 Using Eq.~\eqref{eqapp:bubble} and Eq.~\eqref{eqapp:bubble_cc}, we have obtained the expression for bosonic self-energy, $\Pi(q)$,  at large gap limit, which is shown in Eq.~\eqref{eq:boson_self_lm1} in the main text. In the small gap limit ($q_0 \ll \delta \ll v_F \vert \bm{q} \vert $), $\Pi(q)$ develops a Landau damping term,
 \begin{equation}
\Pi(q)=\bar{g}\nu_F \left[ 1-\frac{\vert q_0 \vert}{v_F \vert \bm{q} \vert}  \right]\, .
 \label{eqapp:boson_self_lm2}
 \end{equation}

Next, we derive the fermionic self-energy, $\Sigma(k_0)$, in both the large gap and small gap limit. We reproduce Eq.~\eqref{eq:fermion_self1} in the main text as follows.
\begin{equation}
 \begin{aligned}
&\Sigma(k_0) =\bar{g}a^2\iint G^d(k-q)D(q)  \, \frac{d^2q}{(2\pi)^2} \, \frac{dq_0}{2\pi} \\
&= -\frac{\bar{g}a^2}{(2\pi)^3} \int dq_0 \int dq \frac{q  D(q)}{v_F \vert \bm{q} \vert} \left(  \int     \frac{d\theta }{\cos\theta- \frac{i(k_0-q_0+i\delta)}{v_F \vert \bm{q} \vert}   }\right) \\
&= -\frac{i\bar{g}a^2}{(2\pi)^2} \int dq_0 \int dq      \left(\frac{q  D(q) \, \text{sgn}(k_0-q_0)}{\sqrt{\left(v_F \vert \bm{q} \vert\right)^2 + \left(k_0-q_0+i\delta \right)^2}} \right) \,.
 \label{eqapp:self_energy1}
 \end{aligned}
\end{equation}
 Here, we have decomposed the integration over $2D$ momentum as
$\int d^2q \rightarrow \int dq \, q \int d\theta$ . 
Now, we investigate Eq.~\eqref{eqapp:self_energy1} in both the large gap and small gap limit. The results are discussed in Section.~\ref{sec3}. In the limit $\delta \gg v_F \vert \bm{q} \vert \gg \vert k_0-q_0 \vert$, Eq.~\eqref{eqapp:self_energy1} reduces to
\begin{equation}
 \begin{aligned}
 \Sigma(k_0) &\approx\frac{-i\bar{g}}{(2\pi)^2} \int dq_0 \int dq  \left( \frac{q  D(q) \, \text{sgn}(k_0-q_0)}{\sqrt{    \left(i\delta \right)^2 +2 i \delta \vert k_0-q_0 \vert \sigma(k_0-q_0) }}           \right) \\
&\approx \frac{-\bar{g}}{(2\pi)^2 } \int dq_0 \int dq \frac{q  D(q) }{\delta} \left(   1- \frac{ \vert k_0-q_0 \vert \, \text{sgn}(k_0-q_0)}{i\delta}    \right) \\
 &=\frac{-\bar{g}}{(2\pi)^2 \delta }  \int dq_0 \int dq \, \frac{q}{\vert \bm{q} \vert^2+\frac{q_0^2}{\tilde{c}^2}} \\
 &=\frac{-\bar{g}}{(2\pi)^2 \delta } \frac{1}{2} \int dq_0 \log(1+\frac{\Lambda^2 \tilde{c}^2}{q_0^2})  \\
 &\approx\frac{-\bar{g}}{4\pi \delta } \Lambda \tilde{c} \, .
  \label{eqapp:self_energy_lm1}
 \end{aligned}
\end{equation}
This expression is obtained using a UV momentum cutoff,  $\Lambda$. We have used here the expression of bosonic propagator $D(q)$ in the large gap limit shown in Eq.~\eqref{eq:boson_propagator_lm1} in the main text. Eq.~\eqref{eqapp:self_energy_lm1} indicates that the self-energy becomes constant and thus renormalizes the chemical potential.
On the other hand, in the small gap limit ($v_F \vert \bm{q} \vert \gg \delta \gg \vert k_0-q_0 \vert$), bosonic propagator assumes the form shown in Eq.~\eqref{eq:boson_propagator_lm2}. In this limit, Eq.~\eqref{eqapp:self_energy1} reduces to a FL form, 
\begin{equation}
 \begin{aligned}
 \Sigma(k_0)  & \approx \frac{-i\bar{g}}{(2\pi)^2 v_F} \int dq_0 \, \text{sgn}(k_0-q_0) \int_{\delta/v_F}^{\Lambda} dq \frac{1}{\vert \bm{q} \vert^2+\frac{g \gamma_F \vert q_0 \vert}{v_F \vert \bm{q} \vert}}  \\
  & \approx \frac{-i\bar{g}}{(2\pi)^2 v_F} \int dq_0  \, \text{sgn}(k_0-q_0) \int_{\delta/v_F}^{\Lambda} dq \frac{1}{\vert \bm{q} \vert^2} \\
   & \approx \frac{-i\bar{g}}{(2\pi)^2 v_F} \frac{v_F}{\delta} \int dq_0 \, \text{sgn}(k_0-q_0) \\
   &= - \frac{i\bar{g}}{2\pi^2 } \frac{k_0}{\delta} \, .
  \label{eqapp:self_energy_lm2}
 \end{aligned}
\end{equation}

\section{\label{app2} Pairing in the large gap limit }
In this appendix, beginning with the self-consistent expression for the pairing vertex in Eq.~\eqref{eq:vertex1}, we derive the gap equation in the large gap limit ($\delta \gg v_F \vert \bm{q}_1 \vert \gg \vert q_{1,0} \vert$ and $\delta \gg v_F \vert \bm{q}_2 \vert \gg \vert q_{2,0} \vert$), as shown in Eq.~\eqref{eqmn:vertex1_freq1_lm1} of the main text.
Below, we present the derivation in two stages. First, we integrate out momentum $q_2$, which leads to
\begin{equation}
 \begin{aligned}
&  a^2\int\frac{d^2q_2}{(2\pi)^2} G^c(-k-q_1+q_2)  G^c(k+q_1-q_2)D(q_2) \\
&=a^2\int\frac{dq_2}{(2\pi)^2}\frac{q_2 D(q_2)}{\left(v_F \vert \bm{q}_2 \vert \right)^2} \int  \frac{d\theta_2}{\frac{\vert \tilde{\Sigma}(-k_0-q_{1,0}+q_{2,0}) \vert^2}{\left( v_F \vert \bm{q}_2 \vert \right)^2}+ \cos^2\theta_2} \\
&= \frac{1}{2\pi v_F} \frac{1}{\vert \tilde{\Sigma}(-k_0-q_{1,0}+q_{2,0}) \vert}   \int \frac{dq_2}{\vert \bm{q}_2\vert^2+\frac{q_{2,0}^2}{\tilde{c}^2}} \\
&= \frac{1}{4 v_F}  \frac{\tilde{c} }{\vert q_{2,0} \vert} \frac{1}{\vert \tilde{\Sigma}(-k_0-q_{1,0}+q_{2,0}) \vert}\,.
 \label{eqapp:vertex_int1_lm1}
 \end{aligned}
\end{equation}
Second, the momentum $q_1$ is integrated out as 
\begin{equation}
 \begin{aligned}
& a^2 \int\frac{d^2q_1}{(2\pi)^2} G^d(k+q_1)  G^d(-k-q_1)D(q_1) \\
 &=a^2\int\frac{dq_1}{(2\pi)^2} \frac{q_1 D(q_1)}{\left( v_F \vert \bm{q}_1 \vert\right)^2}  \int   \frac{d\theta_1}{\frac{\vert \tilde{\Sigma}(k_0+q_{1,0}) \vert^2}{\left( v_F \vert \bm{q}_1 \vert \right)^2}+ \left(\cos\theta_1+\frac{\delta}{v_F \vert \bm{q}_1 \vert} \right)^2} \\
   &=a^2\int\frac{dq_1}{(2\pi)^2} \frac{\pi q_1 D(q_1) }{\vert \tilde{\Sigma}(k_0+q_{1,0}) \vert} \left[ \frac{1}{\sqrt{\left(v_F \vert \bm{q}_1 \vert \right)^2 +\left( \vert \tilde{\Sigma}(k_0+q_{1,0}) \vert +i\delta  \right)^2}} + \frac{1}{\sqrt{ \left( v_F \vert \bm{q}_1 \vert \right)^2+\left( \vert \tilde{\Sigma}(k_0+q_{1,0}) \vert -i\delta \right)^2}} \right]  \\
   &\approx a^2\int\frac{dq_1}{(2\pi)^2}  \frac{q_1 D(q_1)\pi }{\vert \tilde{\Sigma}(k_0+q_{1,0}) \vert} \left[\frac{1}{i\delta \sqrt{ 1+\frac{2 \tilde{\Sigma}(k_0+q_{1,0})} {i\delta}    }}+     \frac{1}{-i\delta \sqrt{ 1-\frac{2 \tilde{\Sigma}(k_0+q_{1,0})} {i\delta} }}   \right]\\
    &\approx a^2\int\frac{dq_1}{(2\pi)^2} \, q_1 D(q_1) \frac{\pi }{\vert \tilde{\Sigma}(k_0+q_{1,0}) \vert} \left[       \frac{1}{i\delta} \left(1- \frac{ \tilde{\Sigma}(k_0+q_{1,0})}{i\delta} \right)     -   \frac{1}{i\delta} \left(1+ \frac{ \tilde{\Sigma}(k_0+q_{1,0})}{i\delta} \right)             \right]\\
    &=\frac{1}{2\pi \delta^2} \int_0^\Lambda dq_1 \frac{q_1}{\vert \bm{q}_1\vert^2+\frac{q_{1,0}^2}{\tilde{c}^2}} \\
    &=\frac{1}{4\pi \delta^2} \log{\left(1+\frac{\Lambda^2 \tilde{c}^2}{q_{1,0}^2}  \right)}\, .
 \label{eqapp:vertex_int2_lm1}
 \end{aligned}
\end{equation}
In Eqs.~\eqref{eqapp:vertex_int1_lm1} and ~\eqref{eqapp:vertex_int2_lm1}, we have defined $\tilde{\Sigma}(k_0)=k_0+i\Sigma(k_0)$. Here, the self-energy $\Sigma(k_0)$ only has a static contribution in the large gap limit, as discussed in Section~\ref {sec3}. We have used the bosonic propagators, $D(q_1)$ and $D(q_2)$, in the large gap limit from   Eq.~\eqref{eq:boson_propagator_lm1} in the main text. In the third step of the Eq.~\eqref{eqapp:vertex_int2_lm1}, we have used the approximation $\delta \gg v_F \vert \bm{q}_1 \vert \gg \vert k_0+ q_{1,0} \vert$.  Using these two expressions in Eq.~\eqref{eq:vertex1} yields
\begin{equation}
 \begin{aligned}
\phi(k_0)=\lambda T^2 \sum_{q_{1,0} \ne 0}\sum_{\substack{q_{2,0} \ne 0 \\\vert q_{2,0} \vert \ne {k_0+q_{1,0}}}} \frac{\phi (k_0+q_{1,0}-q_{2,0})}{\vert \tilde{\Sigma}(-k_0-q_{1,0}+q_{2,0}) \vert}  \frac{1 }{\vert q_{2,0} \vert} \log{\left(1+\frac{\omega_\Lambda^2}{q_{1,0}^2}\right)}.
 \label{eqapp:vertex1_freq_lm1}
 \end{aligned}
\end{equation}
where the effective coupling constant is given by $\lambda=\bar{g}^2  \tilde{c}  / 16\pi v_F \delta^2 $, and we have defined $\omega_\Lambda=\Lambda \tilde{c}$.
We can redefine frequencies $k_0  \rightarrow \tilde\omega_{m}$, $ k_0+q_{1,0}-q_{2,0} \rightarrow \tilde\omega_{m^\prime}$, $k_0+q_{1,0} \rightarrow \tilde\omega_{m^{\prime \prime}}$ and rewrite Eq.~\eqref{eqapp:vertex1_freq_lm1} as 
\begin{equation}
 \begin{aligned}
\phi(\tilde\omega_m)= \lambda T^2 \sum_{m^\prime \ne m^{\prime \prime} }\sum_{  m^{\prime \prime} \ne m  }  \frac{\phi (\tilde\omega_{m^\prime})}{\vert \tilde\omega_{m^\prime} \vert}  \frac{1 }{\vert \tilde\omega_{m^{\prime \prime}} -\tilde\omega_m{^\prime} \vert} \log{\left(1+\frac{\omega_\Lambda^2}{   \left( \tilde\omega_{m^{\prime \prime}} -\tilde\omega_m  \right)^2 }\right)}\,,
 \label{eqapp:vertex1_freq1_lm1}
 \end{aligned}
\end{equation}
This is Eq.~\eqref{eqmn:vertex1_freq1_lm1} in the main text.  
The numerical solution of this gap equation reveals a parametric enhancement of $T_c$ over the BCS result, as shown in Fig.~\ref{tc_num_soln}. 
We have already discussed the details of the numerical technique to compute $T_c$ in Section~\ref{sec4_1}.
To elucidate the origin of this enhancement, we now simplify Eq.~\eqref{eqapp:vertex1_freq_lm1} analytically.  
We start by converting the frequency summations in this expression to integrations, $T\sum_{q_{1,0}} \rightarrow \int dq_{1,0}/2\pi$ and $T\sum_{q_{2,0}}\rightarrow \int dq_{2,0}/2\pi$, followed by a rescaling of integration variables as $q_{1,0}/\omega_\Lambda \rightarrow q_{1,0}$ and $q_{2,0}/\omega_\Lambda \rightarrow q_{2,0}$. Integrating out the frequencies (with $k_0 \approx 0$) leads to
\begin{equation}
 \begin{aligned} 
1&=\frac{\lambda}{2\pi^2}\int \left(\int \frac{1 }{\vert q_{2,0} \vert} \left(  \frac{1}{\vert q_{1,0}+q_{2,0} \vert} +  \frac{1}{\vert q_{1,0}-q_{2,0} \vert}   \right) dq_{2,0} \right)  \log{\left(1+\frac{1}{q_{1,0}^2}\right)} dq_{1,0}\\
& \approx \frac{2\lambda}{\pi^2}\int_{T_c/{\omega_\Lambda}}^1 \left(\frac{1}{\vert q_{1,0}  \vert} \int_{T_c/{\omega_\Lambda}}^{ \vert q_{1,0} \vert }      \frac{1}{\vert q_{2,0}\vert }  dq_{2,0}  +    \int_{\vert q_{1,0}  \vert}^1 \frac{1 }{\vert q_{2,0} \vert^2 }    dq_{2,0} \right)  \log{\left(\frac{1}{q_{1,0}}\right)} dq_{1,0} \\
& \approx \frac{2\lambda}{\pi^2} \int_{T_c/{\omega_\Lambda}}^1 \left(\frac{1}{q_{1,0}}  \log{\left(\frac{q_{1,0}}{T_c/{\omega_\Lambda}}\right)}+ \frac{1}{q_{1,0}}     \right)  \log{\left(\frac{1}{q_{1,0}}\right)} dq_{1,0} \\
& = \frac{2\lambda}{\pi^2} \int_{T_c/{\omega_\Lambda}}^1  \left(  - \frac{1}{q_{1,0}}   \left( \log{\left(\frac{1}{q_{1,0}} \right)} \right)^2          - \frac{1}{q_{1,0}}  \log{\left(\frac{1}{q_{1,0}} \right)} \log{\left(\frac{T_c}{{\omega_\Lambda}}\right)}    + \frac{1}{q_{1,0}}  \log{\left(\frac{1}{q_{1,0}} \right)}             \right)dq_{1,0} \\
& = \frac{\lambda}{2\pi^2} \left( -\frac{2}{3}   \left(\log{\left(\frac{T_c}{{\omega_\Lambda}}\right)} \right)^3  +2 \left(\log{\left(\frac{T_c}{{\omega_\Lambda}}\right)} \right)^2  \right) \, .
 \label{eqapp:vertex1_freqsum1_lm1}
 \end{aligned}
\end{equation}
The terms with cubic and squared logarithm of $T_c$ in this expression guarantee a larger $T_c$ than the BCS expression resulting from a term linear in logarithmic of $T_c$. The solution of this expression is shown in Fig.~\ref{tc_num_soln}(b) along with the numerical solution.

\begin{figure*}[t]
\includegraphics[width=0.8\hsize]{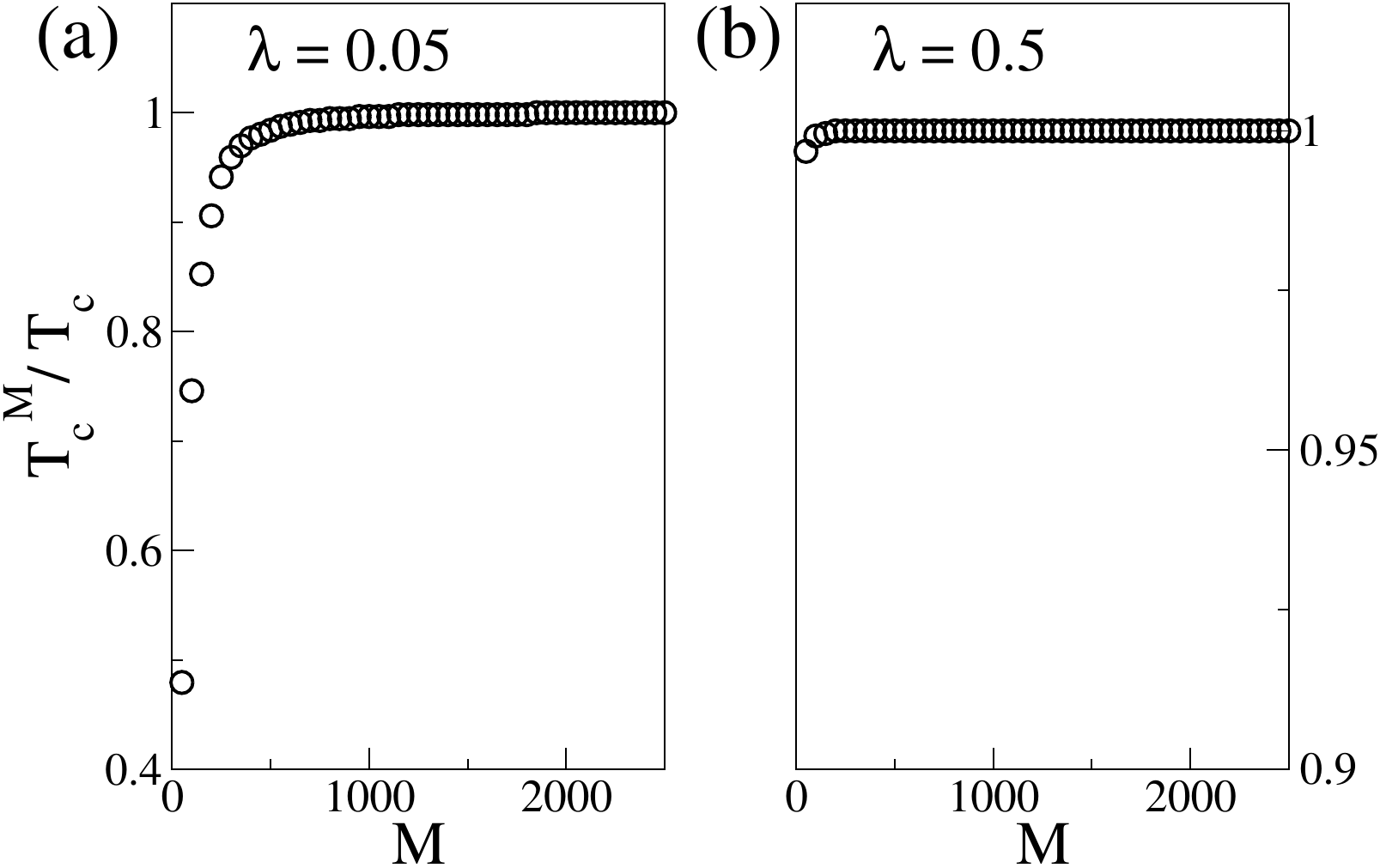}
\caption{\label{tc_cutoff} The critical temperature $T_c^M$, evaluated as a function of frequency cutoff $M$, is normalized by its asymptotic value $T_c$, obtained at large $M$ where it becomes independent of $M$. 
Panels (a) and (b) correspond to the coupling $\lambda=0.05$ ($T_c=0.00056 \, \omega_\Lambda$) and $0.5$ ($T_c=0.025 \, \omega_\Lambda$), respectively. 
Notably, for $\lambda=0.05$, $T_c^M$ converges to $T_c$ at $M\ge 800$, while for $\lambda=0.5$, the convergence is already achieved at $M\ge 200$.
}
\end{figure*}

As emphasized in Section~\ref{sec4_1}, a sufficiently large frequency cutoff $M$ in Eq.~\eqref{eqmn:vertex1_freq1_lm1} is essential for an accurate numerical evaluation of $T_c$. We now show that, for a smaller cutoff, the value of the critical temperature becomes sensitive to the chosen cutoff. We compute the critical temperature for various cutoffs and denote it as $T_c^M$, while the value obtained at a sufficiently large cutoff, where the solution no longer depends on $M$, is denoted by $T_c$. Fig.~\ref{tc_cutoff} shows the ratio $T_c^M/T_c$ as a function of $M$ for two different coupling strength. The result demonstrates that a small cutoff is sufficient at a strong coupling, whereas a larger cutoff is crucial at smaller coupling. The choice $M=2000$ ensures an accurate estimation of $T_c$ across the entire range of coupling $\lambda$ considered in Fig.~\ref{tc_num_soln}.
%


\section{\label{app4} Pairing in the small gap limit }
Now, we reproduce the gap equation in the small gap limit ($v_F \vert \bm{q}_1 \vert \gg \delta \gg \vert q_{1,0} \vert$ and $v_F \vert \bm{q}_2 \vert \gg \delta \gg \vert q_{2,0} \vert$), which is given in Eq.~\eqref{eq:vertex1_freq_lm2} in the main text. Integrating out momentum $q_2$ in  Eq.~\eqref{eq:vertex1} yields
\begin{equation}
 \begin{aligned}
& a^2 \int\frac{d^2q_2}{(2\pi)^2} G^c(-k-q_1+q_2)  G^c(k+q_1-q_2)D(q_2) \\
&=\frac{a^2}{2\pi v_F} \frac{1}{\vert \tilde{\Sigma}(-k_0-q_{1,0}+q_{2,0}) \vert}   \int_{\delta/v_F}^\Lambda dq_2 \frac{1}{\vert \bm{q}_2\vert^2 a^2+\frac{g\gamma_F \vert q_{2,0}\vert}{v_F \vert \bm{q}_2 \vert}} \\
& \approx \frac{1}{2\pi v_F} \frac{1}{\vert \tilde{\Sigma}(-k_0-q_{1,0}+q_{2,0}) \vert}   \int_{\delta/v_F}^\Lambda dq_2 \frac{1}{\vert \bm{q}_2\vert^2} \\
&= \frac{1}{2\pi \delta}   \frac{1}{\vert \tilde{\Sigma}(-k_0-q_{1,0}+q_{2,0}) \vert}\,.
 \label{eqapp:vertex_int1_lm2}
 \end{aligned}
\end{equation}
Similarly, we integrate out momentum $q_1$ in  Eq.~\eqref{eq:vertex1} as 
\begin{equation}
 \begin{aligned}
&a^2 \int\frac{d^2q_1}{(2\pi)^2} G^d(k+q_1)  G^d(-k-q_1)D(q_1) \\
  &=a^2\int\frac{dq_1}{(2\pi)^2} \, q_1 D(q_1)  \frac{\pi }{v_F \vert \bm{q}_1 \vert \vert \tilde{\Sigma}(k_0+q_{1,0}) \vert} \left[ \frac{1}{\sqrt{1+\left( \frac{\vert \tilde{\Sigma}(k_0+q_{1,0}) \vert +i\delta}{v_F \vert \bm{q}_1 \vert} \right)^2}} + \frac{1}{\sqrt{1+\left( \frac{\vert \tilde{\Sigma}(k_0+q_{1,0}) \vert -i\delta}{v_F \vert \bm{q}_1 \vert} \right)^2}}  \right]  \\
  &\approx  \frac{1}{2\pi v_F} \frac{1 }{\vert \tilde{\Sigma}(k_0+q_{1,0}) \vert}   \int_{\delta/v_F}^\Lambda dq_1 \frac{1}{\vert \bm{q}_1\vert^2}   \\
&= \frac{1}{2\pi \delta}   \frac{1}{\vert \tilde{\Sigma}(-k_0-q_{1,0}) \vert}   \,.
 \label{eqapp:vertex_int2_lm2}
 \end{aligned}
\end{equation}
Notably, in these two expressions, the IR momentum cutoff is set by $\delta$, which allows us to neglect the Landau damping in $D(q_1)$ and $D(q_2)$.
  Using Eqs.~\eqref{eqapp:vertex_int1_lm2} and ~\eqref{eqapp:vertex_int1_lm2} in Eq.~\eqref{eq:vertex1}, we obtain
\begin{equation}
 \begin{aligned}
\phi(k_0) &=\frac{\bar{g}^2  T^2}{4\pi \delta^2} \sum_{\substack{q_{1,0}; \\ \vert q_{1,0} \vert \ne \vert k_0 \vert   }}\sum_{\substack{q_{2,0}; \\ \vert q_{2,0} \vert \ne \\ \vert k_0+q_{1,0} \vert   }} \phi (k_0+q_{1,0}-q_{2,0}) \frac{1}{\vert \tilde{\Sigma}(-k_0-q_{1,0}+q_{2,0}) \vert}  \frac{1}{\vert \tilde{\Sigma}(-k_0-q_{1,0}) \vert} \\
&=\frac{\bar{g}^2  T^2}{4\pi \delta^2} \frac{1}{\left(1+\frac{ \bar{g}a^2}{2\pi^2 \delta}\right)^2} \sum_{\substack{q_{1,0}; \\ \vert q_{1,0} \vert \ne \vert k_0 \vert   }}\sum_{\substack{q_{2,0}; \\ \vert q_{2,0} \vert \ne \\ \vert k_0+q_{1,0} \vert   }} \phi (k_0+q_{1,0}-q_{2,0}) \frac{1}{\vert k_0+q_{1,0}-q_{2,0} \vert}  \frac{1}{\vert k_0+q_{1,0} \vert}.
 \label{eqapp:vertex1_freq_lm2}
 \end{aligned}
\end{equation}
Thus, we obtain Eq.~\eqref{eq:vertex1_freq_lm2} in the main text. Here, we have used the FL form of fermionic self-energy $\Sigma$, as discussed in Section~\ref{sec3} and given in Eq.~\eqref{eqapp:self_energy_lm2}. The effective coupling constant is given by 
\begin{equation}
 \begin{aligned}
 \lambda_{s}=\frac{\bar{g}^2 /4\pi \delta^2}{ \left(1+\frac{ \bar{g}a^2}{2\pi^2 \delta} \right)^2} \approx \frac{1}{4\pi}\left(\frac{\bar{g}}{\delta}\right)^2\, .
  \label{eqapp:coupling_lm2}
  \end{aligned}
\end{equation}
Next, we obtain Eq.~\eqref{eq:vertex1_freqsum2_lm2} in the main text by integrating the frequencies from $T_c$ to $\Lambda_\Omega$ (with $k_0 \approx 0$) as shown below.
\begin{equation}
 \begin{aligned}
1&=\frac{2\lambda_s}{(2\pi)^2}\int_{T_c}^{\Lambda_\Omega} \left(\int_{T_c}^{\Lambda_\Omega} \frac{1 }{\vert q_{1,0} \vert} \left(  \frac{1}{\vert q_{1,0}+q_{2,0} \vert} +  \frac{1}{\vert q_{1,0}-q_{2,0} \vert}   \right) dq_{2,0} \right) dq_{1,0}\\
& =\frac{\lambda_s}{2\pi^2}\int_{T_c}^{\Lambda_\Omega} \left(\frac{2}{\vert q_{1,0}  \vert^2} \int_{T_c}^{\vert q_{1,0}  \vert}     dq_{2,0}  + \frac{2}{\vert q_{1,0}  \vert}   \int_{\vert q_{1,0}  \vert}^{\Lambda_\Omega} \frac{1 }{\vert q_{2,0} \vert}        dq_{2,0} \right)dq_{1,0} \\
&= \frac{\lambda_s}{\pi^2}\int_{T_c}^{\Lambda_\Omega} \left(\frac{1}{\vert q_{1,0}  \vert} -   \frac{T_c}{\vert q_{1,0}  \vert^2}+ \frac{1}{\vert q_{1,0}  \vert} \log{\frac{\Lambda_\Omega}{\vert q_{1,0}  \vert}} \right) dq_{1,0} \\
  &\approx\frac{\lambda_s}{\pi^2}  \left[ \log{\frac{\Lambda_\Omega}{T_c}} +\frac{1}{2} \left(  \log{\frac{\Lambda_\Omega}{T_c}} \right)^2   \right] \, .
 \label{eqapp:vertex1_freqsum1_lm2}
 \end{aligned}
\end{equation}

\twocolumngrid
\bibliography{ferroelectric_ref}

\end{document}